\documentclass[aps,prl,amsmath,amssymb,nofootinbib,twocolumn]{revtex4}
\usepackage{lipsum}
\usepackage{amsfonts}
\usepackage{graphicx}
\usepackage{float}
\usepackage{mathptmx} 
\usepackage{appendix}
\begin{document}

\title{Coherent control of optical bistability in Rydberg electromagnetically-induced transparency atomic system}

\author{You-Lin Chuang}\email{yloptics@cts.nthu.edu.tw}
\affiliation{Physics Division, National Center for Theoretical Sciences, Hsinchu 30013, Taiwan}
\author{Mehmet Emre Tasgin}
\affiliation{Hacettepe University, Institute of Nuclear Sciences, 06800, Ankara, Turkey}

\begin{abstract}

We study optical bistable behavior of a Rydberg electromagnetically-induced transparency (EIT) atomic medium in a unidirectional optical ring-cavity. Due to strong van del Waal (vdW) interactions between the atoms, both optical nonlinear dispersion and nonlinear absorption coefficients are  enhanced substantially. Under the condition of two-photon on resonance, we show that probe one-photon detuning can change the phase of the third order nonlinearity coefficient, which tunes the character of the optical bistability within different ratios of dispersive and absorptive types. This enables the single-photon control over photonic devices for further manipulation of light other than switches and transistors. More interestingly, we predict appearance of a scaling phenomena for optical bistabilities with the factors of coupling Rabi frequency and atomic density. 
Additionally, we also discuss the influence of the cavity detuning and the mirror transmission coefficient on the optical bistable behavior. The strong bistable feature provides a good ingredient for realizing all-optical logic gate devices in optical computing.    
\end{abstract}

\date{\today}
\maketitle%

\section{i. introduction}

Quantum coherence and interference play important roles in atomic and  optical systems. 
In the past decades, the phenomenon of electromagnetically induced transparency (EIT)~\cite{EIT}, which arises due to quantum interference between two or more transition paths with atom-field interactions, has been investigated extensively.
In the presence of quantum coherence, some optical properties can be modified, e.g. enhancement of Kerr nonlinearity~\cite{giantKerr, LargeKerr}, large refraction index~\cite{EnhanceN1, EnhanceN2, EnhanceN3}, and negative refraction index~\cite{NegativeN}. 
Consequently, many applications based on coherent media, such as large cross-phase modulation \cite{XPM}, photon switching~\cite{photon_switching}, four-wave mixing~\cite{FWM}, laser without inversion~\cite{LWI}, spectral narrowing \cite{spectral_narrowing}, slow light~\cite{SL1, SL2} and light storage~\cite{LS}, are widely studied.

Although coherent media possess the rich applications mentioned above, 
the nonlinearity of the media is still too weak so that it is difficult to observe significant nonlinear optical effects at low-intensity illumination regime.  
We can provide the follwoing comparison for the optical nonlinearities, i.e. Kerr nonlinear coefficient $ \chi^{(3)} $, in different atomic systems. In conventional three-level atomic system, nonlinear coefficient assigns a small value $ \chi^{(3)} \sim 10^{-15} (\text{m}^2 \text{V}^2) $. A giant Kerr nonlinear coefficient can be efficiently obtained based on EIT-scheme \cite{giantKerr}, which was demonstrated on the atomic ensemble of sodium by L. V. Hau {\it et al}.~\cite{SL1}. The measured Kerr coefficient is $ \chi^{(3)} \sim 7\times10^{-8} (\text{m}^2 \text{V}^2) $, which corresponds to an about 8 orders of magnitude enhancement. Apart from the usual atomic systems, Rydberg atoms, which has an excited state with a high principle quantum number, provide huge polarizability~\cite{Rydberg1, Rydberg2}. Strong interactions between the Rydberg atoms give an alternative mechanism for increasing the nonlinear response~\cite{NLRyEIT}. When Rydberg atoms are combined with the EIT effect, a Rydberg-EIT system, the system has quantum interference effect with long-lived coherence as well as strong long-range atom-atom interactions arise from dipole-dipole or van del Waals (vdW) interactions. This shifts the Rydberg level and results in Rydberg blockade when the interaction-induced shift is much larger than the EIT linewidth.
Thus, in a Rydberg-EIT medium, one obtains a strong optical Kerr nonlinearity $ \chi^{(3)} \sim 5\times10^{-2} (\text{m}^2 \text{V}^2) $, which is strong enough to enable quantum nonlinear optics in single-photon level. In addition, such a strong nonlinearity also provides a direct evidence for the existence of Rydberg blockade sphere~\cite{blockade1, blockade2, blockade3, NLRyEIT}. 

Besides the enhancement of nonllinearity, an accompanying phenomenon, called optical bistability (OB) or optical multistability (OM), has been studied in various systems, e.g., semiconductor quantum well structures~\cite{SC,SQW1,SQW2,QW}, silicon ring resonators~\cite{ring_resonators1, ring_resonators2}, and silicon waveguide-resonators~\cite{waveguide}. 
OB is characterized by a hysteresis curve which depends on the history, and exhibits rapid jumps between two stable states, of lower and of higher outputs, for the same input. This requires a nonlinear system with a feedback provided by the cavity. Optical cavities containing OB and OM atomic systems have been investigated both experimentally~\cite{OBexp1, OBexp2, OBexp3} and theoretically~\cite{theoOB1, theoOB2, theoOB3, theoOB4, theoOB5}. Depending on the type of the nonlinear response of a medium, i.e. nonlinear absorption or nonlinear dispersion, it is possible to classify different media, as absorptive and dispersive bistability, respectively~\cite{NL_book}.  

All-optical switch and transistor, in which the gate light pulse can change the transmission of a target light pulse even in single-photon level~\cite{Single_photon_switching1, Single_photon_switching2, Single_photon_switching3}, have already been demonstrated in Rydberg atomic systems via utilizing the Rydberg blockade effect. In difference to these achievements, extinguishing the target light in the presence of the gate light pulse, in this work we tune the nonlinearity feature of the OB at desired rates of dispersion and absorption. Hence, we have different transmissions of the probe field in our system by modifying the feature of the OB via using different incident probe field intensities. The hysteresis, which depends on the history of the path, can be used in photonic logic gate devices, playing an important role in the development of optical computing. Single-photon control of the OB feature enables the further manipulation of light in photonic devices.

In this paper, we study a Rydberg EIT medium in a unidirectional ring cavity possessing optical bistabilities, via strong vdW interactions between the Rydberg atoms, even for a weak probe field input. We show that phase of the nonlinear coefficient can be manipulated by one-photon detuning of the probe, which can change the bistable feature of the medium dramatically. Coupling (Rabi) frequency and atomic density are also two important factors which show scaling properties on the bistable curves. We also discuss the influences of the cavity detuning and the transmission coefficient of the mirror. We observe that magnitude of the nonlinear coefficient is directly proportional to the volume of the Rydberg blockage sphere, which means that appearance of the optical bistable response is supported by the Rydberg blockage nonlinearity mechanism.

The paper is organized as follows. In Sec. II, we describe the theoretical model for an optical bistable system. We derive the equations of motion revealing the Rydberg EIT features. By solving the intra-field equations of motion, we obtain an analytical formula for the transmission of the probe field, which shows a dispersive-type bistability.  
In Sec. III, we demonstrate our results for different physical parameters, discuss the underlying physics. Finally, we present our conclusions in Sec. IV.


\section{ii. system and theoretical model}
\label{sec2}

In this section, we consider the system which is depicted in Fig. 1. A Rydberg EIT medium is placed inside a unidirectional ring cavity system, which consists of two mirrors with reflection $ R $ and transmission $ T $ ($ R + T = 1 $), and the other two mirrors with $ 100\% $ reflectivity, as shown in Fig. 1(a). Rydberg EIT system is characterized by a cascade three-level atomic configuration, in which the two fields, probe and coupling field, are interacting with the atomic system with the corresponding Rabi frequency $ \Omega_p $ and $ \Omega_c $, respectively, as shown in Fig. 1(b). $ \Delta_p = \omega_p - \omega_{21} $ and $ \Delta_c = \omega_c - \omega_{32} $ are the one-photon detunings of probe and coupling fields, where $ \omega_{\mu\nu}\equiv (E_\mu - E_\nu)/\hbar $ is the frequency difference between state $ \vert\mu\rangle $ and $ \vert\nu\rangle $.
According to the atom-field interactions, we can derive the equations of motion for the first order atomic operators under EIT condition, i.e. $ \Omega_p \ll \Omega_c $.
\begin{eqnarray}
&&\dfrac{\partial}{\partial t}\hat{\sigma}_{12} = -\left( \gamma_{12}-i\Delta_p\right) \hat{\sigma}_{12} + \dfrac{i}{2}\Omega_p + \dfrac{i}{2}\Omega_c^{\ast}\hat{\sigma}_{13},\label{s12}\\
&&\dfrac{\partial}{\partial t}\hat{\sigma}_{13} = -\left( \gamma_{13}-i\Delta_2\right) \hat{\sigma}_{13} + \dfrac{i}{2}\Omega_c\hat{\sigma}_{12}-i\hat{U}\hat{\sigma}_{13}, \label{s13}
\end{eqnarray}
where $ \hat{\sigma}_{\mu\nu} \equiv \vert\mu\rangle\langle\nu\vert $ is the atomic flip operator. 
$ \Delta_2 \equiv \Delta_p + \Delta_c $ is the two-photon detuning. 
$ \gamma_{12} $ and $ \gamma_{13} $ are the relaxation rates of $ \hat{\sigma}_{12} $ and $ \hat{\sigma}_{13} $, and the corresponding Langevin noise terms are ignored at this stage.
Here the total vdW induced shift in Eq.(\ref{s13}) is defined as $ \hat{U} \equiv \int d^3\textbf{r}' V(\textbf{r}-\textbf{r}')n(\textbf{r}')\hat{\sigma}_{33}(\textbf{r}') $, in which $ n $ is atomic number density of Rydberg medium, and $ V = C_6/\vert \textbf{r}_i - \textbf{r}_j\vert^6 $ is the frequency shift of a pair of atoms located at $ \textbf{r}_i $ and $ \textbf{r}_j $.

\begin{figure}
\begin{center}
\includegraphics[scale=0.19]{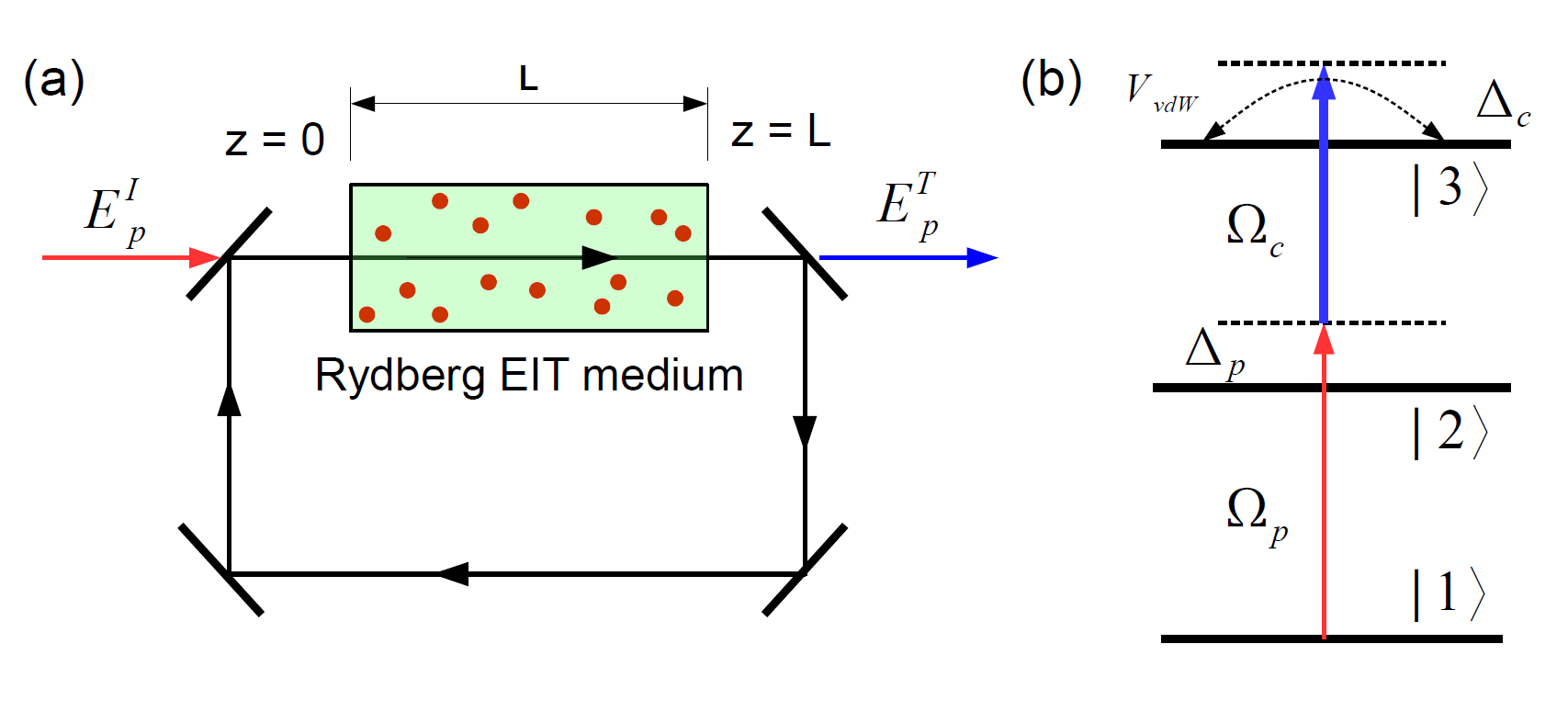} 
\end{center}
\caption{ (a) an unidirectional ring cavity system in which a Rydberg EIT medium is contained. (b) The cascade-type configuration of Rydberg EIT atom.}
\end{figure}

By solving Eq.(\ref{s12}) and (\ref{s13}) in steady-state regime and letting two-photon on resonance $ \Delta_2 = 0 $, we can obtain the expectation value of $ \hat{\sigma}_{12} $ defined as $ \rho_{21}\equiv\langle\hat{\sigma}_{12}\rangle $, which is proportional to the dipole transition of probe field.
\begin{eqnarray}
\label{rho}
\rho_{21} = \dfrac{2i\gamma_{13}}{\Omega^2}\Omega_p - \dfrac{\Omega_p\vert\Omega_c\vert^4}{\Omega^2\vert\Omega\vert^4}n\int d^3 \textbf{r}'\dfrac{2V(\textbf{r}-\textbf{r}')\vert\Omega_p(\textbf{r}')\vert^2}{\Omega^2 +2i\Gamma V(\textbf{r}-\textbf{r}')},\nonumber\\
\end{eqnarray}
in which $ \Gamma\equiv \Gamma_2-2i\Delta_p $, where $ \Gamma_2\equiv2\gamma_{12} $ is the spontaneous emission of state $ \vert 2\rangle $. $ \Omega^2 \equiv \vert\Omega_c\vert^2+2\Gamma\gamma_{13} $. The detail derivations of Eq.~(\ref{rho}) are given in Appendix A.

In Eq.(\ref{rho}), we have two parts which are linear part in first term, and the nonlinear term in second term. Since $ \gamma_{13}\vert\Omega_p\vert \ll \vert\Omega_c\vert^2 $ is satisfied in EIT condition, we can ignore the linear term safely, and only take the nonlinear term into account.
With the dipole source term coming from Rydberg EIT medium, we can obtain the propagation equation of probe field as shown as follows. 
\begin{eqnarray}
\label{FE}
\dfrac{\partial \Omega_p}{\partial z} = in\mu^2\dfrac{k}{\epsilon\hbar}\rho_{21}
\end{eqnarray}
where $ \mu $ is the dipole moment from $ \vert 1\rangle $ and $ \vert 2\rangle $. $ k $ and $ \epsilon $ are the wavevector and electric permittivity at the transition frequency $ \omega_{21} $, respectively.
Together with Eq.~(\ref{rho},~\ref{FE}) and associating with standard field propagation equation, we have
\begin{eqnarray}
\label{kerrFE}
\dfrac{\partial E_p}{\partial \zeta} = i\dfrac{kL}{2}\chi^{(3)}\Gamma_2^2 \vert E_p\vert^2 E_p = 
i\eta\vert E_p\vert^2 E_p, 
\end{eqnarray}
where $ E_p \equiv \Omega_p/\Gamma_2 $ is a dimensionless field amplitude. $ \zeta\equiv z/L $ is a dimensionless length, where $ L $ is the medium length.
To obtain Eq.(\ref{kerrFE}), we have assumed that $ \Omega^2 \simeq \vert\Omega_c\vert^2 $, and 
$ \vert\Omega_p(\textbf{r}')\vert^2 \simeq \vert\Omega_p(\textbf{r})\vert^2 $ \cite{NLRyEIT} so that we can move it outside from the integral of Eq.(\ref{rho}).
After calculating the integral, we can obtain an analytical form of the dimensionless nonlinear coefficient $ \eta $ which is given as (please see the detail derivations in Appendix A)
\begin{eqnarray}
\label{ita}
\eta\simeq \dfrac{\pi\alpha}{2\sqrt{2}}\left( \dfrac{\Gamma_2}{\vert\Omega_c\vert}\right)^2 \left( \dfrac{4}{3}\pi R_c^3 n\right)\left( -1+i\right) \left(1-2i\dfrac{\Delta_p}{\Gamma_2} \right)^{-1/2} , \nonumber\\
\end{eqnarray}
in which $ \alpha = n \sigma_{\text{abs}} L $ is the optical density of Rydberg EIT medium, where $  \sigma_{\text{abs}} = 3\lambda^2/2\pi $ is the probe absorption cross section.  
$ R_c \equiv \left( C_6/\delta_{\text{EIT}}\right)^{1/6} $ is the radius of Rydberg blockade sphere, and $ \delta_{\text{EIT}}\equiv \vert\Omega_c\vert^2/\gamma_{12} $ corresponds to the width of EIT window. The physical meaning of blockade radius $ R_c $ is defined by the distance at which the vdW induced frequency shift is larger than $  \delta_{\text{EIT}} $. 
From Eq.(\ref{ita}), we can clearly to see that the nonlinear term $ \eta \vert E_p\vert^2 $ is proportional to the ratio of $ \vert\Omega_p\vert^2/\vert\Omega_c\vert^2 $, but can be greatly enhanced by the value of $ (4\pi/3) R_c^3 n $ \cite{NLRyEIT}, which is the atom numbers inside a Rydberg blockade sphere.  
Besides, $ \eta $ is a complex, and the real and imaginary part correspond to the dispersion and absorption effect of probe field, which can be controlled by tuning the probe one-photon detuning $ \Delta_p $.

Next, we consider the feedback process via the ring-cavity system. 
The input and transmitted probe field are represented by $ E_p^I $ and $ E_p^T $, respectively. The relation between input and output and intracavity fields are known as boundary conditions \cite{theoOB4} given by
\begin{eqnarray}
&& E_p^T = \sqrt{T} E_p(\zeta = 1), \label{bc1}\\
&& E_p(0) = \sqrt{T} E_p^I + Re^{-i\delta}E_p(\zeta = 1)\label{bc2}
\end{eqnarray}
where $ T $ is the transmission coefficient, and the reflection coefficient is defined by $ R $, which satisfy $ T + R = 1 $.
$ \delta = (\omega_{\text{cav}}-\omega_p)\Lambda/c $ is the cavity detuning, and $ \omega_{\text{cav}} $ is the frequency of cavity nearest to resonance, and $ \Lambda $ is the total optical path of the cavity. We are interested in the relationship between input and output fields. As a result, we can rewrite Eq.(\ref{bc1}) and (\ref{bc2}) by
\begin{eqnarray}
&& I_t \equiv \vert E_p^T\vert^2 = T \vert E_p(1)\vert^2, \label{BC1}\\
&& I_i \equiv \vert E_p^I\vert^2 = \dfrac{1}{T}\Big\vert E_p(0) - Re^{-i\delta}E_p(1)\Big\vert^2 \label{BC2}
\end{eqnarray}
In Eq.(\ref{BC1}, \ref{BC2}), the input-output fields are determined by intracavity field, which can be obtained by solving Eq.(\ref{kerrFE}).
The nonlinear nature of Rydberg EIT medium with the feedback from ring cavity system provide necessary conditions to form optical bistability.

In order to study the nonlinear mechanism of optical bistability, we  solve Eq.(\ref{kerrFE}) analytically. We give the detail derivations in Appendix B, in which the analytical solution can be obtained as follows.
\begin{eqnarray}
E_p(\zeta) = E_p(0)\left( 1 + 2 \text{Im}(\eta)\vert E_p(0)\vert^2 \zeta\right)^{i\eta/2 \text{Im}(\eta)} 
\label{Epsol}
\end{eqnarray}
In Eq.(\ref{Epsol}), it is clear to see that $ \text{Im}(\eta) $ provides a dissipation effect. In general, one can rewrite this solution as $ E_p(1) = E_p(0) \exp\left( -\Lambda/2 + i\varphi\right)  $, where $ \Lambda \equiv \ln\left( 1 + 2b\vert E_p(0)\vert^2\right)  $ and $ \varphi\equiv a\Lambda/2b $ represent the decay and phase terms, respectively. We have used $ \eta\equiv a + i b $, in which $ a $ and $ b $ correspond to nonlinear dispersion and absorption.

Substituting Eq.(\ref{Epsol}) into Eq.(\ref{BC1}, \ref{BC2}), 
we can calculate the field transmission defined by $ \mathfrak{T} \equiv I_t/I_i $.

\begin{eqnarray}
\mathfrak{T} = \dfrac{T^2 e^{-\Lambda}}{\left( 1-R~e^{-\Lambda/2}\right)^2 + 4R~e^{-\Lambda/2} \sin^2\left[ (\delta-\varphi)/2\right] }
\label{transmission}
\end{eqnarray}

When $ \Lambda = 0 $ (i.e. $ b = 0 $), Eq.(\ref{transmission}) can reduce back to standard transmission of a ring cavity system or Fabry-Perot interferometer \cite{NL_book}. 
It should be noted that the $ \varphi $ can be expressed as the function of transmitted field intensity, as shown as $ \varphi = -(a/2b)\ln\left( 1 - 2bI_t/T\right)  $, and $ \varphi = a I_t/T $ in the absence of $ b $. Thus, from Eq.(\ref{transmission}), one can realize that the system is a dispersive bistable case, but with nonlinear absorption effect.
In the following section, we will show the results under different physical parameters.

\section{iii. results and discussions}

In this section, we present the numerical results for input-output relation by solving Eq.(\ref{kerrFE}, \ref{BC1}, \ref{BC2}) for different corresponding parameters: Rabi frequency of coupling field $ \Omega_c $, probe detuning $ \Delta_p $, cavity detuning $ \delta $, transmission $ T $, and optical density of the atomic medium $ \alpha $, 
which can be tuned independently. 

\begin{figure}[h]
\begin{center}
\includegraphics[scale=0.285]{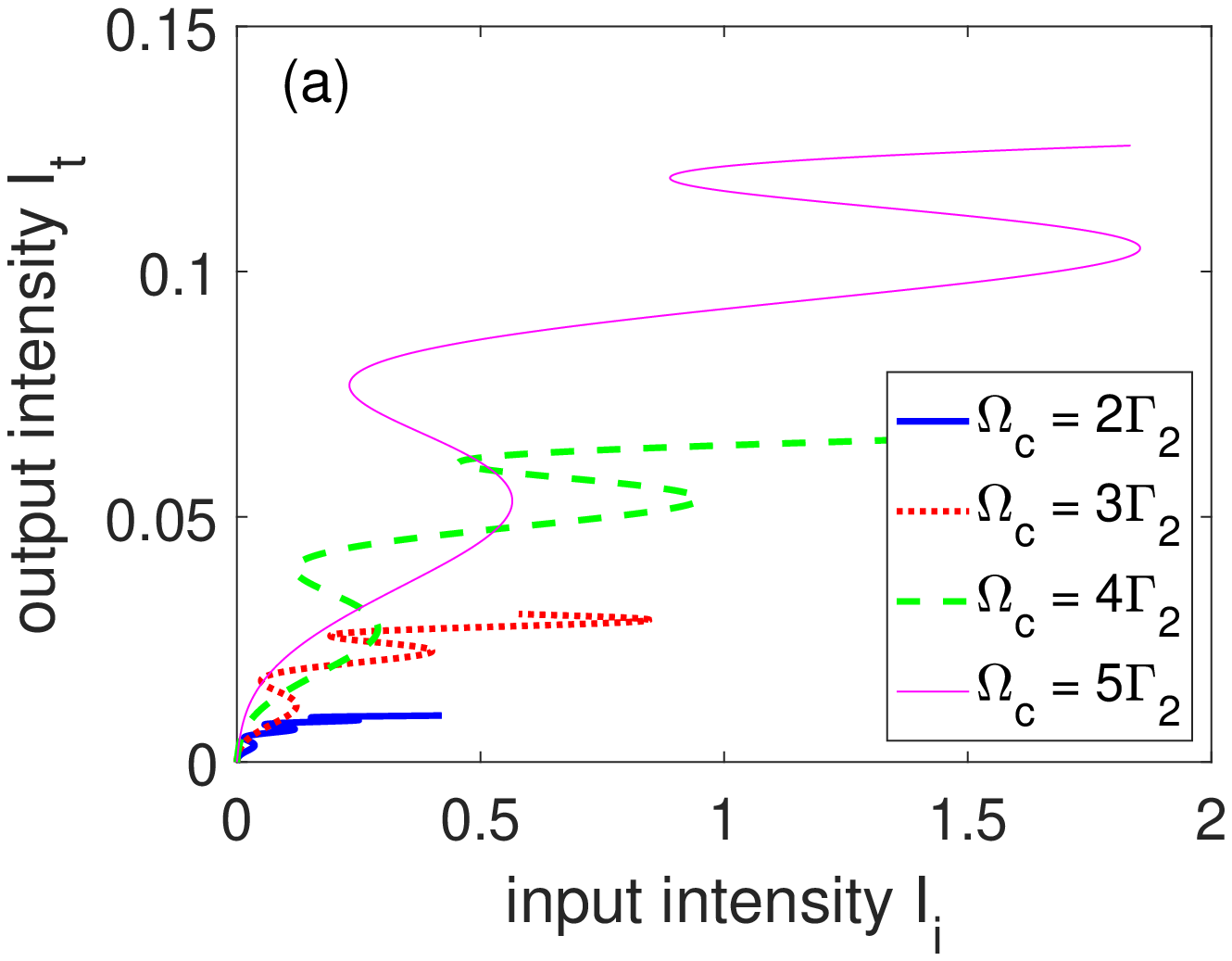} 
\includegraphics[scale=0.285]{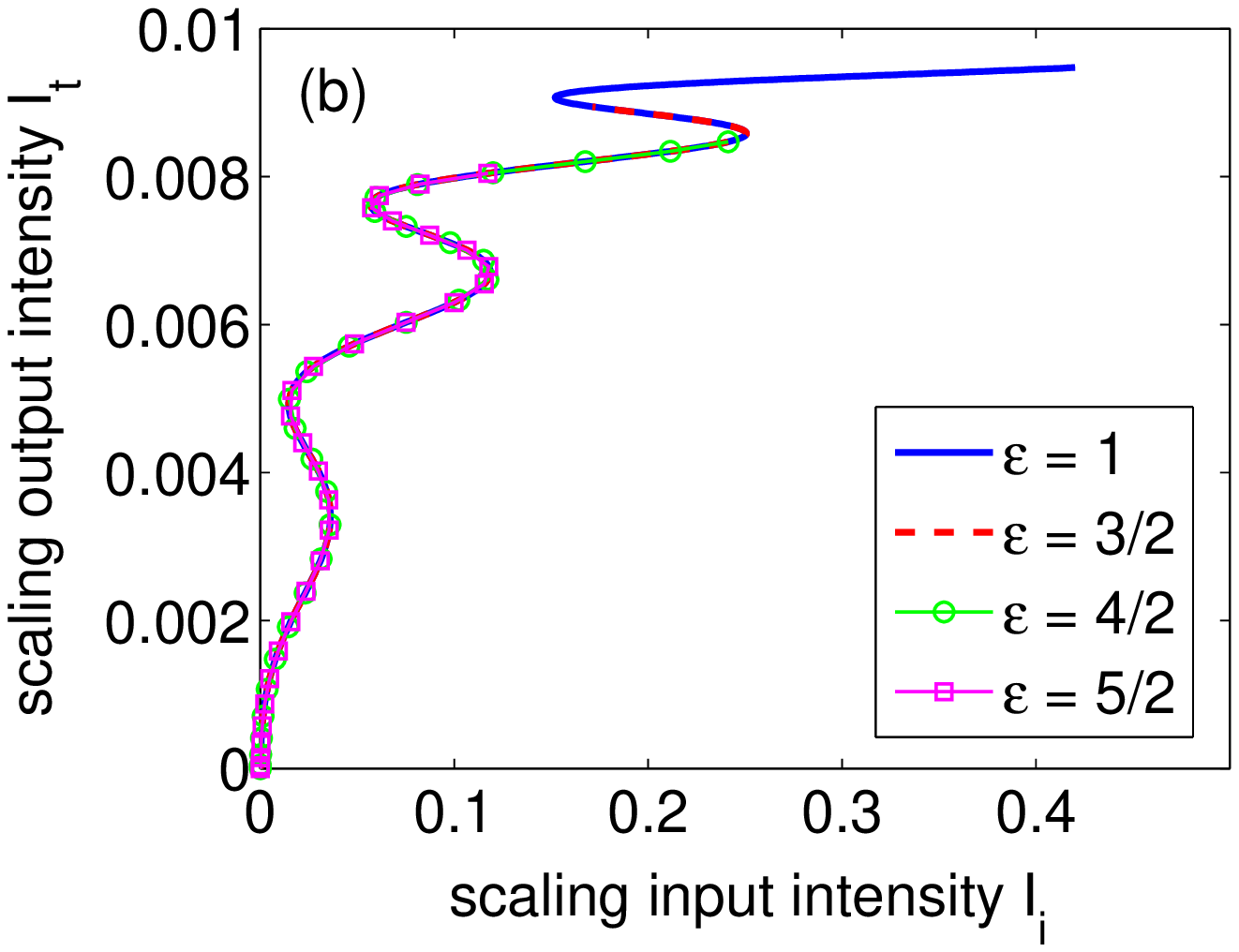} 
\end{center}
\caption{ Optical multi-stabilities with different coupling Rabi frequencies. (a) Blue solid line, red dotted line, green dashed line, and thin magenta line correspond to $ \Omega_c = 2\Gamma_2 $, $ \Omega_c = 3\Gamma_2 $, $ \Omega_c = 4\Gamma_2 $, and $ \Omega_c = 5\Gamma_2 $, respectively. Other parameters are given by $ \Delta_p = 5\Gamma_2 $, $  \delta = 0 $, $ T = 0.5 $, and $ \alpha = 70 $. 
(b) It shows the scaling results with the factor of $ \Omega_c^{-3} $.
By setting the reference curve $ \varepsilon = 1 $ ($ \Omega_c = 2\Gamma_2 $), the input and output fields of other curves are rescaled by the factor $ \varepsilon $.
\label{FigOc}}
\end{figure}

First of all, we study the optical bistable phenomenon with the effect of the Rabi frequency of coupling field. 
As shown in Fig.~\ref{FigOc}(a), we can see the optical bistable signatures between input and output fields. The output field intensity is increasing when $ \Omega_c $ is getting larger, which can be understood with the width of EIT transparent window. From Eq.(\ref{ita}), the nonlinear coefficient $ \eta $ is proportional to $ \vert\Omega_c\vert^{-3} $. 
It means that $ \eta $ decreases when $ \Omega_c $ increases, thus one need to have  
stronger input intensity $ I_i $ to obtain the same transmission.  
As a reason, we rescale the input and output field intensities by multiplying a factor 
$ I_{i,t} \rightarrow \varepsilon^{-3} I_{i,t}$, where $ \varepsilon\equiv \Omega_c/\Omega_{c,0} $. Here $ \Omega_{c,0} $ is the reference Rabi frequency, which is chosen 
$ \Omega_{c,0} = 2\Gamma_2 $ (blue curve) in our case. In Fig.~\ref{FigOc}(b), we can find that the four curves shown in Fig.~\ref{FigOc}(a) fall on the same curve.
It should be noted that the the input probe field intensity is not necessary to be much smaller than that of coupling field. The low-intensity approximation under EIT condition is only used for intra-fields, thus we require $ \Omega_p = 0.2\Omega_c $ in our simulations to guarantee the validity of EIT approximation.  

\begin{figure}
\begin{center}
\includegraphics[scale=0.55]{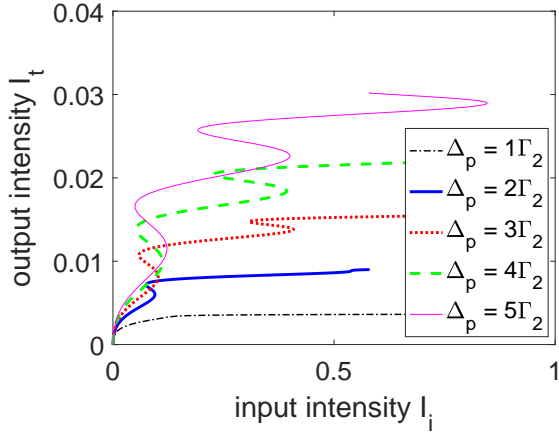} 
\end{center}
\caption{Optical multi- and bistable curves with different probe detunings. The input-output relations are shown by black dash-dotted line ($ \Delta_p = 1\Gamma_2 $), Blue solid line ($ \Delta_p = 2\Gamma_2 $), red dotted line ($ \Delta_p = 3\Gamma_2 $), green dashed line ($ \Delta_p = 4\Gamma_2 $), and thin magenta line ($ \Delta_p = 5\Gamma_2 $). Other parameters are set by $ \Omega_c = 3\Gamma_2 $, $ \delta = 0 $, $ T = 0.5 $, and $ \alpha = 70 $.
\label{FigDp}}
\end{figure}

\begin{figure}
\begin{center}
\includegraphics[scale=0.55]{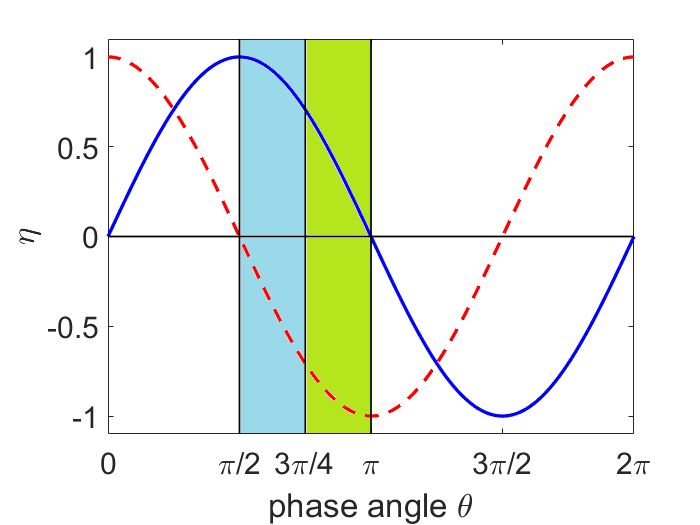} 
\end{center}
\caption{Real and imaginary part of $ \eta $ verse the corresponding phase angle $ \theta $. Red dashed curve and blue solid curve represent the real and imaginary part of $ \eta $, which correspond to dispersive and absorptive term of system. \label{phase}}
\end{figure}

Next, we discuss how the probe detuning affect the behaviour of optical multi- and bistability. In Fig.~\ref{FigDp}, we have shown the relations between input and output probe field intensity under various probe detunings. It is obvious to see that probe detuning plays an important role in optical bistabilities. For larger probe detuning, as one can see, the output intensity increases and more bistabilities appear. When the probe detuning decrease to $ \Delta_p = 1\Gamma_2 $, the signature of bistability completely disappears. 
In order to interpret the physics behind this result, we consider the nonlinear coefficient $ \eta $. 
According to Eq.(\ref{ita}), one can see that the phase of $ \eta $ depends on $ \Delta_p $, which can affects the nonlinear degree. The total phase of $ \eta $ is $ \theta = 3\pi/4 + \phi/2$, where $ \phi\equiv\arctan\left( 2\Delta_p/\Gamma_2\right)  $,  ranging from $ -\pi/2< \phi <\pi/2 $. It implies that $ \pi/2 < \theta < \pi $.
As an illustration in Fig. \ref{phase}, the possible phase angle is on the second octant, in which the first half (blue region) corresponds to $ \Delta_p < 0 $, while the other half (green region) corresponds to $ \Delta_p > 0 $. The blue solid curve represents $ \text{Im}(\eta) $, which corresponds to nonlinear absorption, and the red dashed curve is plotted for $ \text{Re}(\eta) $, corresponding to nonlinear dispersive term. 
From Fig.~\ref{phase}, in the blue region, one can see that the absorption is getting larger while $ \vert\Delta_p\vert $ is increasing, at the same time, the dispersion part is decreasing.  
Thus, the nonlinear response is inhibited so that we can't have optical bistability for $ \Delta_p < 0$. 
In contrast, in green region, the absorption is decreasing while $ \Delta_p $ is increasing. Simultaneously, the dispersive part is getting larger. As a result, we can explain the behaviours shown in Fig.~\ref{FigDp}. The transmitted fields have larger intensity and more bistabilities at larger $ \Delta_p $ due to stronger nonlinear dispersion and lower absorptions.

The bistable curve is characterized the nonlinear response of field, including absorption and dispersion. 
Although the nonlinear absorption is quite large, the dispersion provides a significant factor, resulting in optical bistable phenomenon. As discussed in Sec.~II, we will show that the system containing Rydberg EIT medium in cavity can be a dispersive optical bistable device.
 
\begin{figure}
\begin{center}
\includegraphics[scale=0.55]{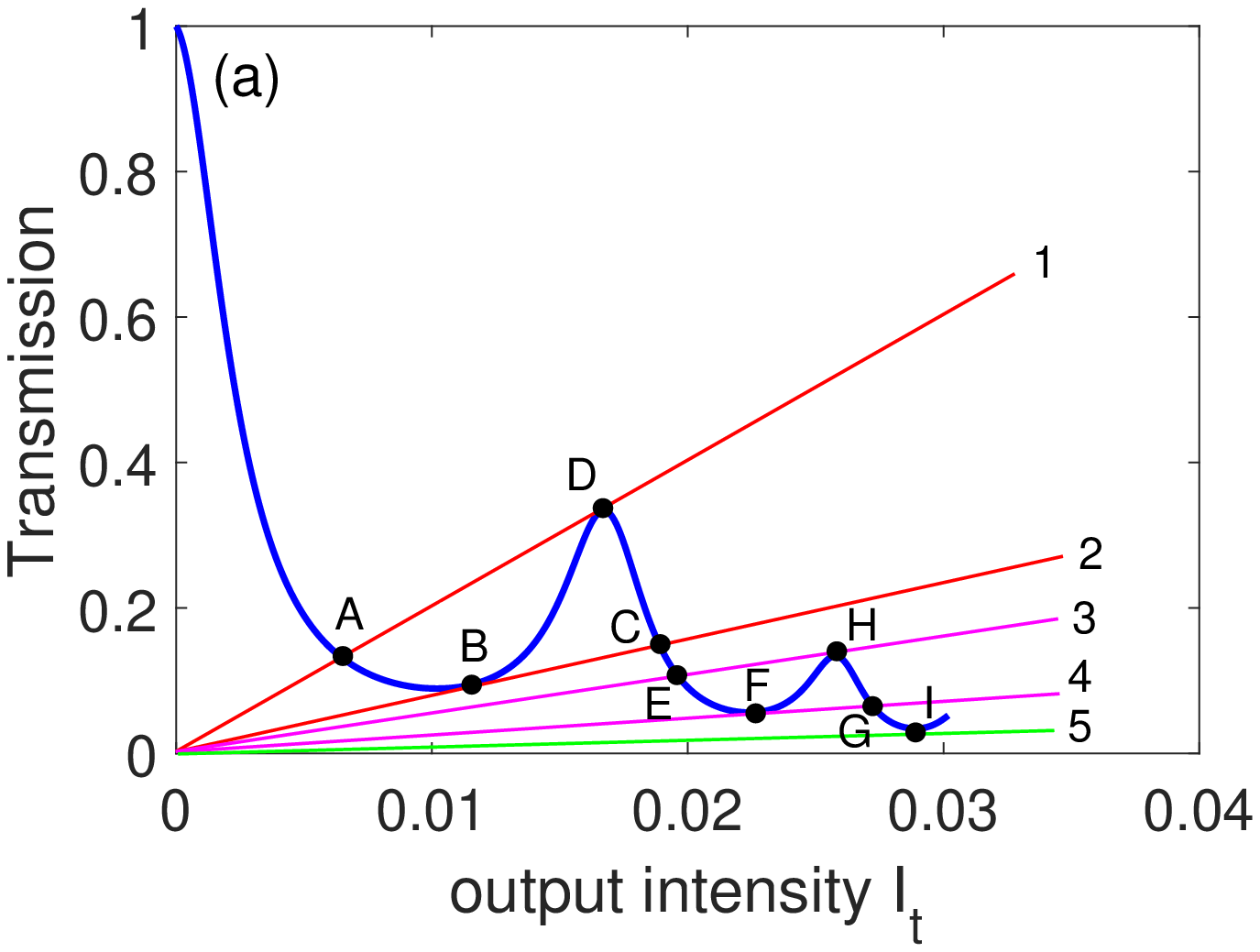} 
\includegraphics[scale=0.55]{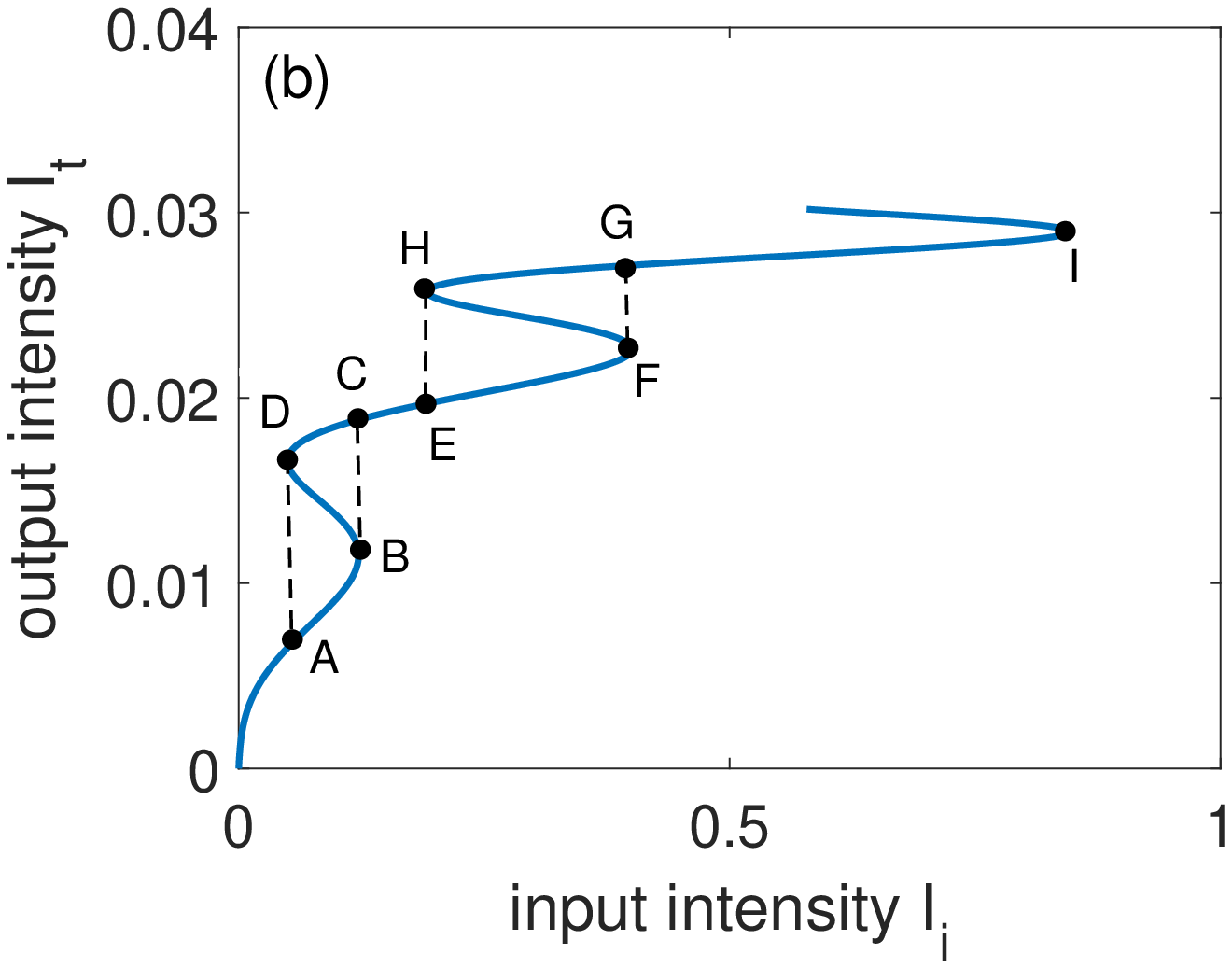} 
\end{center}
\caption{(a) Transmission v.s. output intensity. (b) input-output relation of optical bistability under the same parameters, which are set by $ \Omega_c = 3\Gamma_2 $, $ \Delta_p = 5\Gamma_2 $, $ \delta = 0 $, $ T = 0.5 $, and $ \alpha = 70 $. 
In (a), line 1 to line 5 correspond to the transmissions from lower to higher input intensities. Points A to I are the solutions for the corresponding output field intensities.
\label{Fig5}}
\end{figure} 
  
In Fig.~\ref{Fig5}(a), we have plotted the relation between transmission and output intensity according to Eq.(\ref{transmission}).  
One can clear to see that the magnitude of each transmission peak and the period between two consecutive peaks are gradually decreasing because of the presence of nonlinear absorption. Fig.~\ref{Fig5}(b) shows the input-output relation for the corresponding situation. Im the following, we start to analyze the bistable behaviour. The straight lines in Fig.~\ref{Fig5}(a) represent  the transmissions with different input intensities. With increasing input intensity, the slope of the line becomes smaller ( from line 1 to line 5). The first bistability occurs between line 1 and line 2, as one can see, when input intensity increases (line 1 $ \rightarrow $ line 2), the system response follows the path $ A\rightarrow B \rightarrow C  $, and after then it would follow the other path $ C\rightarrow D \rightarrow A $ when input intensity decreases (line 2 $ \rightarrow $ line 1). Similarly, the second bistability which occurs between line 3 and line 4 follows the loop from $ E\rightarrow F \rightarrow G $ for increasing $ I_i $ (line 3 $ \rightarrow $ line 4), and $ G\rightarrow H \rightarrow E $ for decreasing $ I_i $ (line 4 $ \rightarrow $ line 3). Moreover, if we increase $ I_i $, it will reach to point $ I $, which belongs to the next bistable region. 
From this picture, it is obviously to realize that the bistable dynamical behaviours would be affected by the dissipation, which is coming from nonlinear absorption. 
As a result of degradations of transmission peaks and periods, the jump distance on $ I_t $ of each bistability decreases, while the distance on $ I_i $ between two adjacent turning points increases.
Accordingly, the larger absorption we have, the faster decreasing of each transmission peak we get, which limits the production of bistability at higher input intensity.
It's in turn to give an explanation of the results shown in Fig.~\ref{FigDp}. 

\begin{figure}
\begin{center}
\includegraphics[scale=0.285]{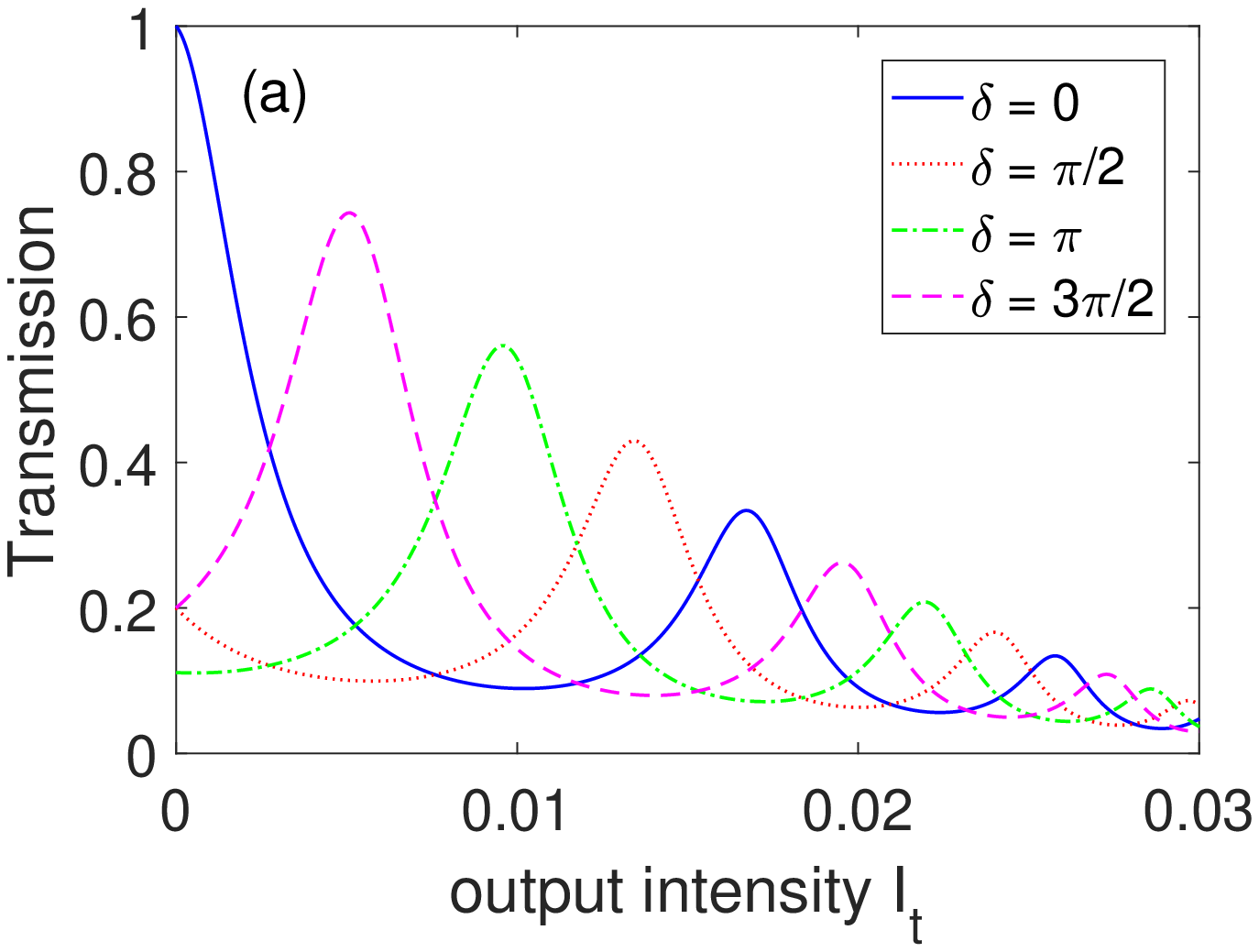} 
\includegraphics[scale=0.285]{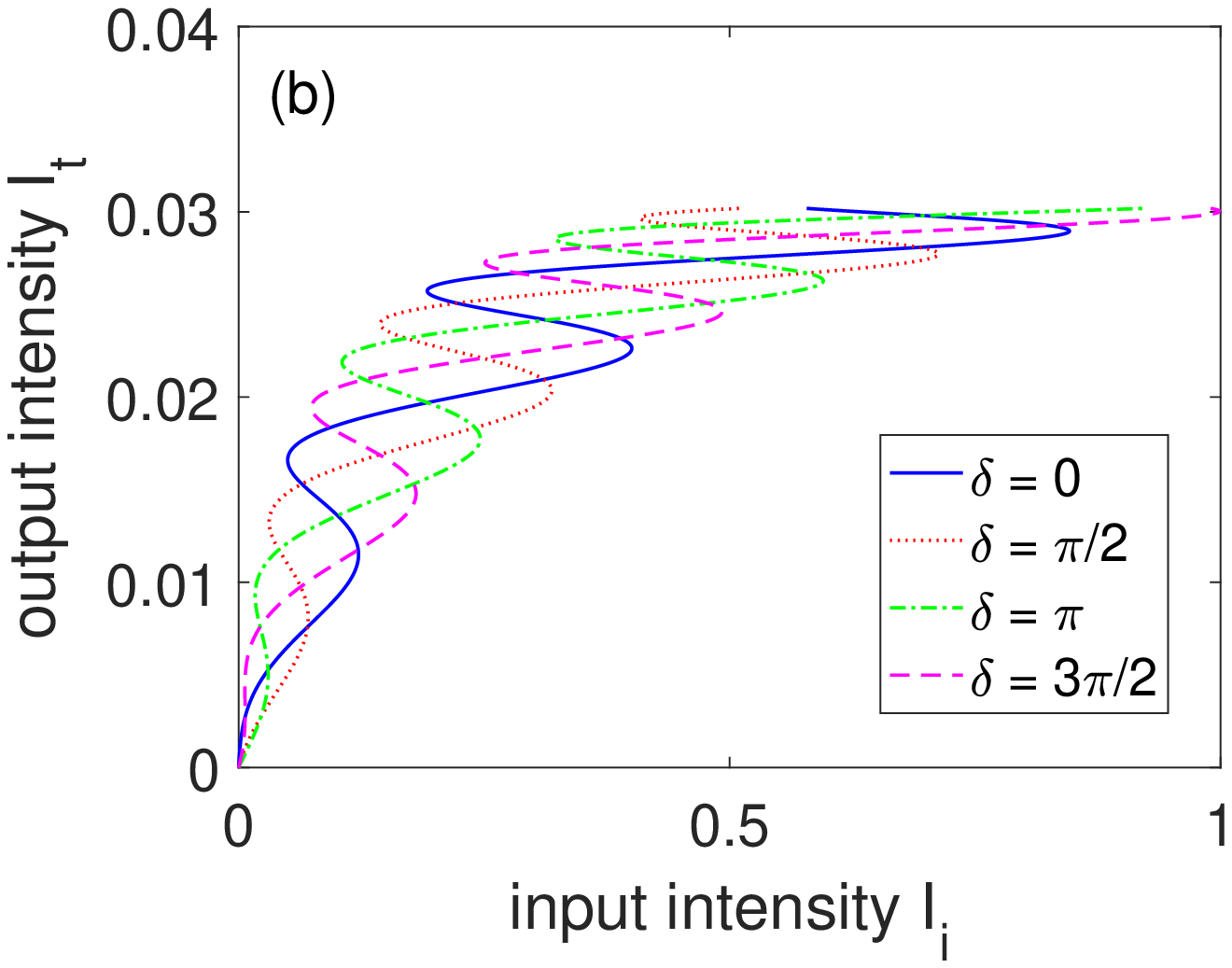} 
\caption{(a) Transmission v.s. output intensity under different cavity detunings. (b) Input-output relation of bistable system with different cavity detunings. Blue solid, red dotted, green dash-dotted, and magenta dashed lines represent the cavity detunings $ \delta = 0,~\pi/2,~\pi $ and $ 3\pi/2 $, respectively. 
\label{FigCD}}
\end{center}
\end{figure}

Then, we consider the factor of cavity detuning $ \delta $. 
In Eq.~(\ref{transmission}), we can see that the cavity detuning effect is just making the shift of transmission peaks, for which the corresponding values would be changed accordingly. 
As one can see in Fig.~\ref{FigCD}(a), the shift of transmission peaks due to cavity detuning is depicted, and the associated input-output relation is shown in Fig.~\ref{FigCD}(b). The blue solid, red dotted, green dash-dotted, and magenta dashed lines represent the cavity detunings $ \delta = 0,~\pi/2,~\pi $ and $ 3\pi/2 $, respectively.  
With the nonlinear absorption, it is evident to see the degradation of transmission peaks with increasing output intensity.   

\begin{figure}[h]
\includegraphics[scale=0.285]{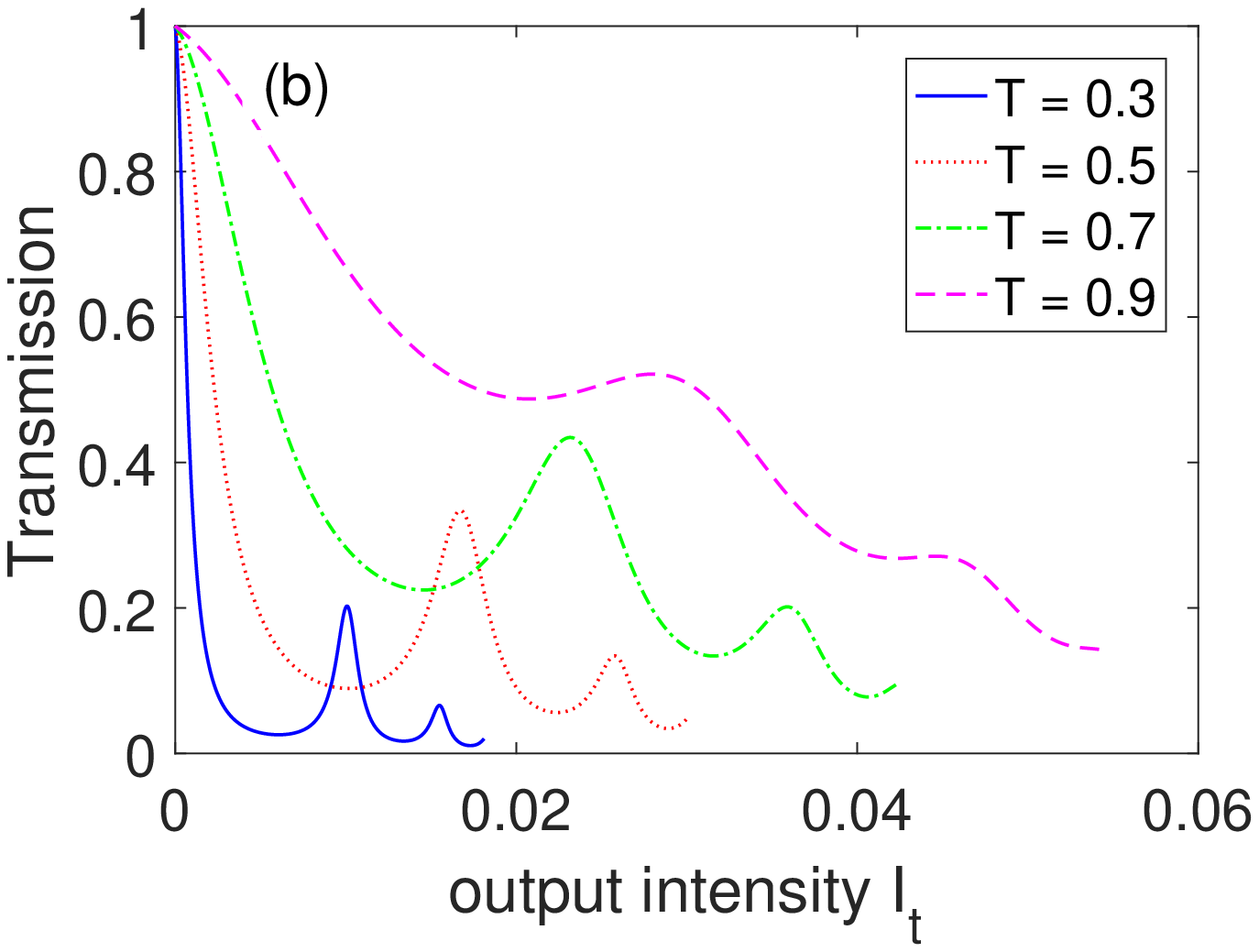} 
\includegraphics[scale=0.285]{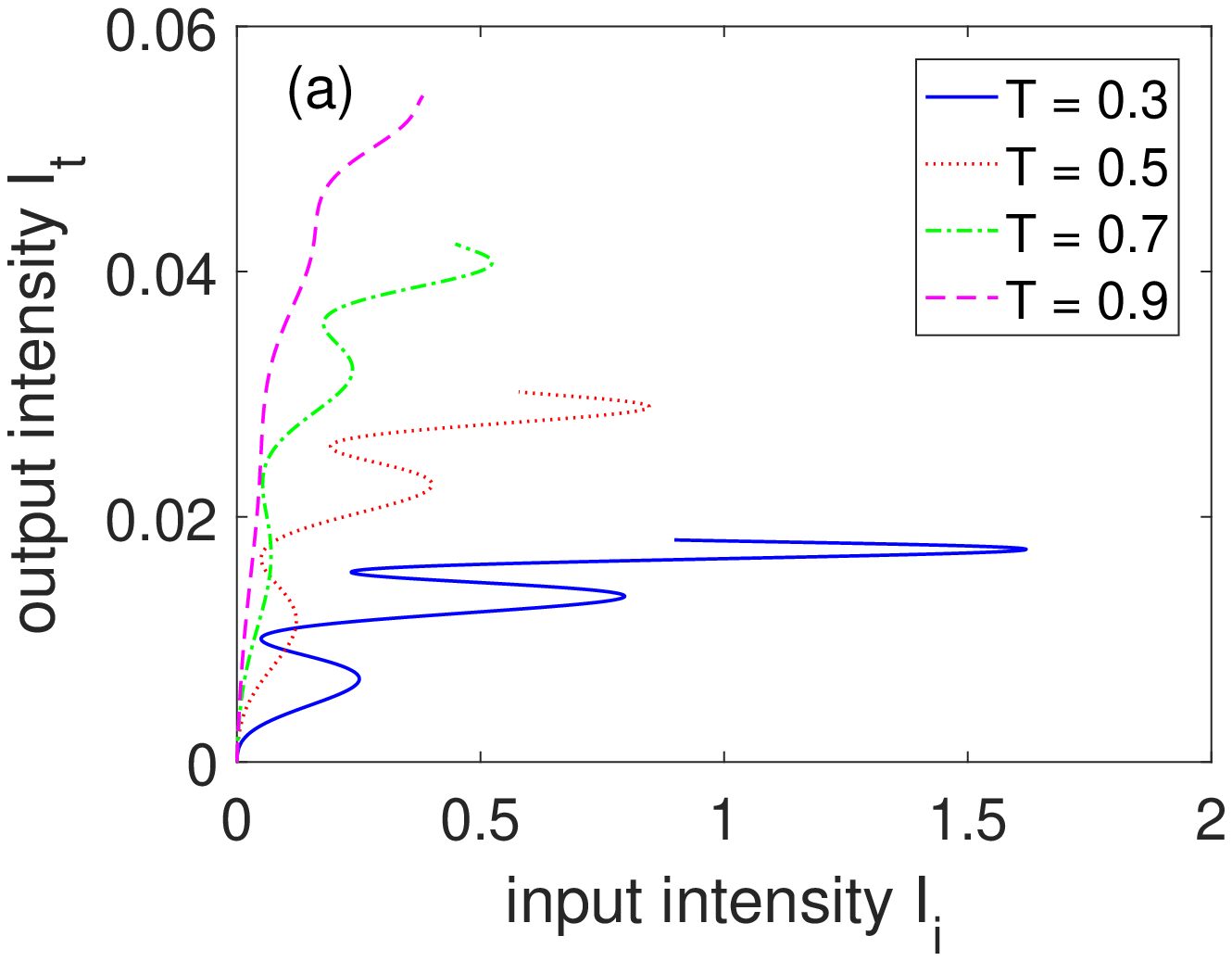} 
\caption{(a) Transmission v.s. output intensity under different mirror transmission coefficients. (b) Input-output relation of bistable system with different mirror transmission coefficients.
Blue solid, red dotted, green dash-dotted, and magenta dashed lines represent the transmission coefficient $ T = 0.3,~0.5,~0.7 $ and $ 0.9 $, respectively. 
\label{FigTM}}
\end{figure}

Using the similar way, we can also study how the transmission coefficient $ T $ affect the bistable properties. 
As shown in Fig.~\ref{FigTM}(a), we can see that the range of $ I_t $ is wider when $ T $ is larger. Besides, the contrasts between   transmission peaks and background become small, which blurs the bistable curve, as the magenta dashed lines shown in Fig.~\ref{FigTM}(b). In contrast, the bistable curve is clear to see when $ T $ is small. Physically, the reason for the fact that the transmission coefficient influences bistable properties is the feedback of the system. According to Eq.(\ref{BC2}), the second term on right hand side is proportional to $ R = 1-T $, which provides the feedback to the system. When $ T $ is getting large, the feedback contribution is decreasing such that the bistable features gradually vanish.

\begin{figure}[h]
\begin{center}
\includegraphics[scale=0.285]{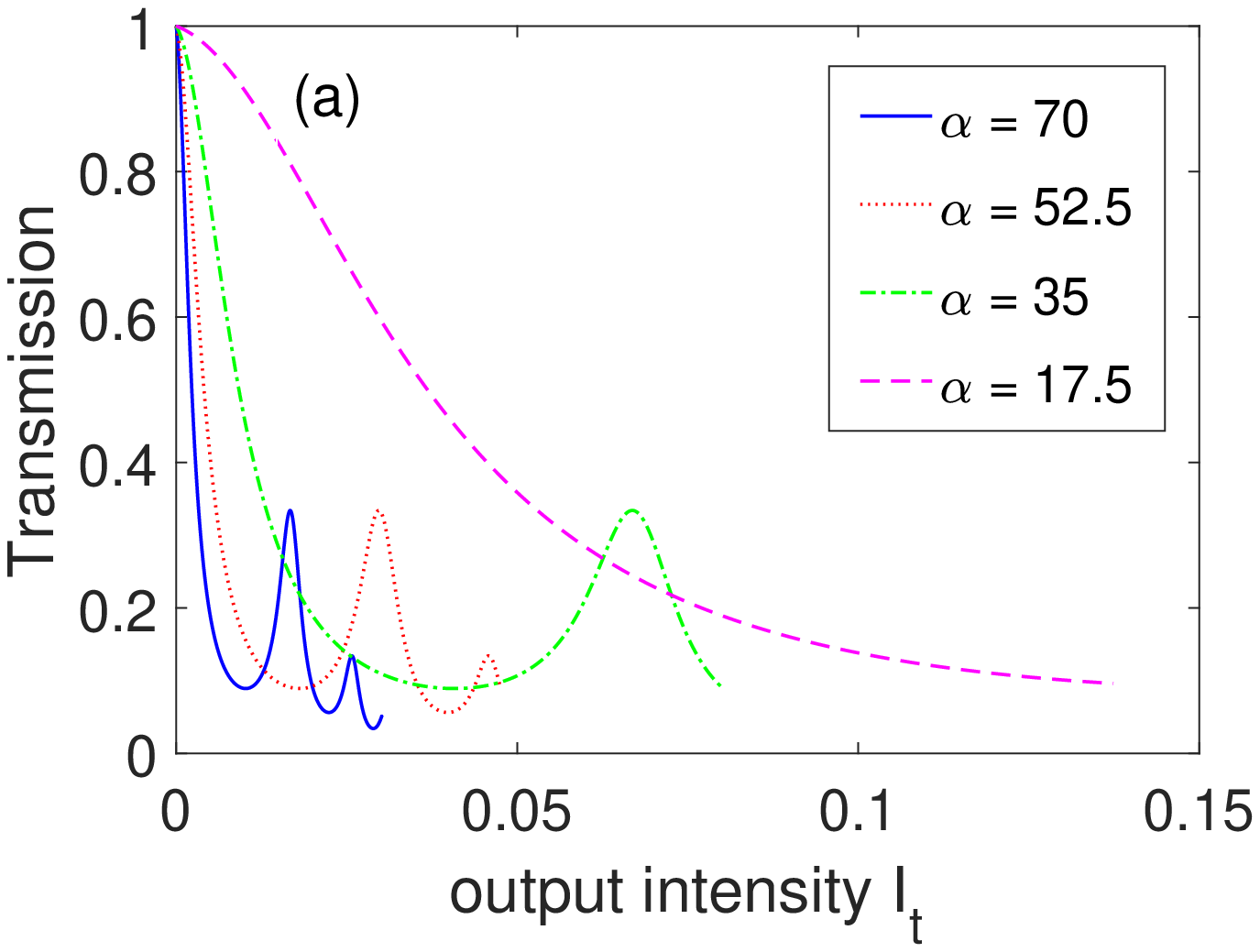} 
\includegraphics[scale=0.285]{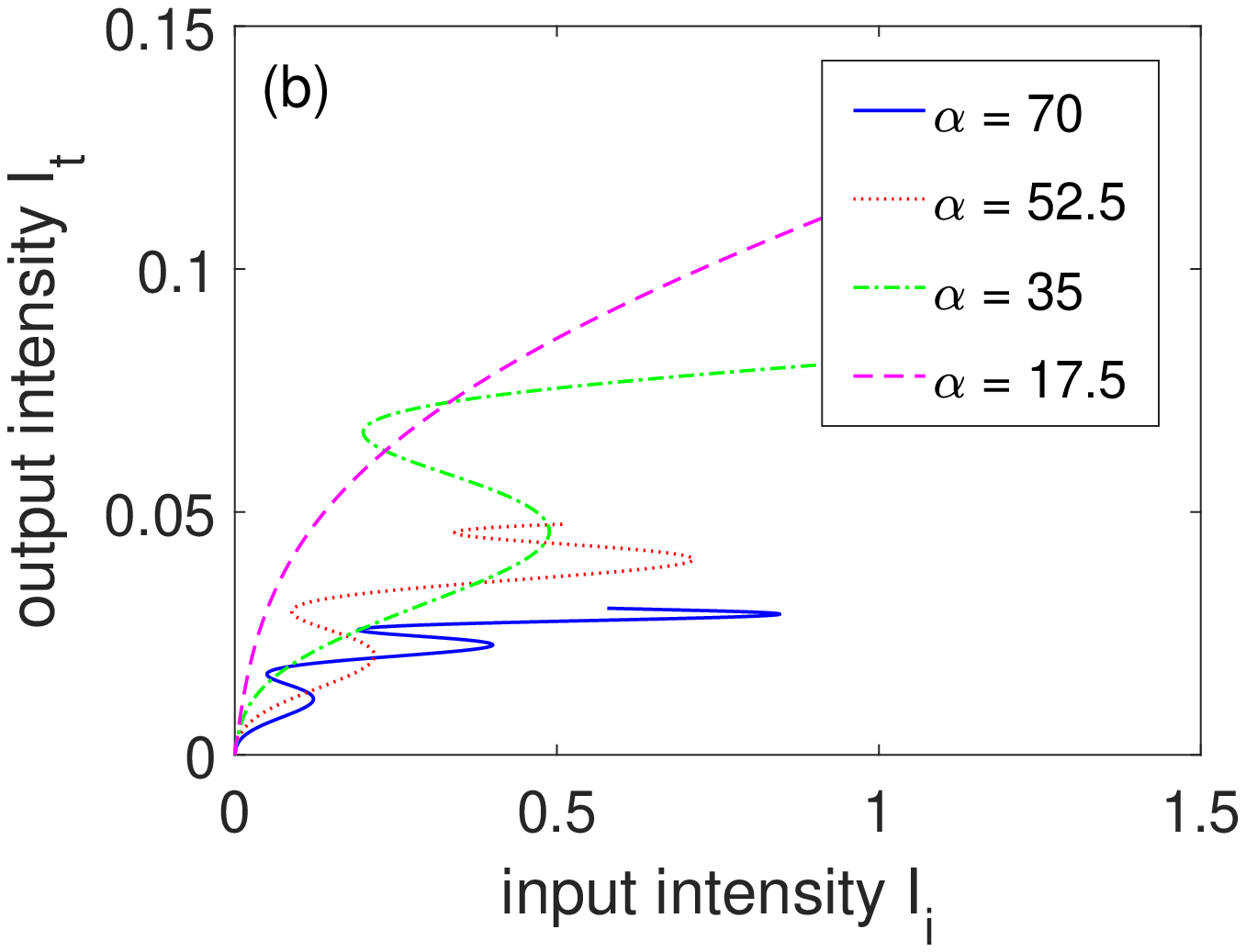} 
\includegraphics[scale=0.285]{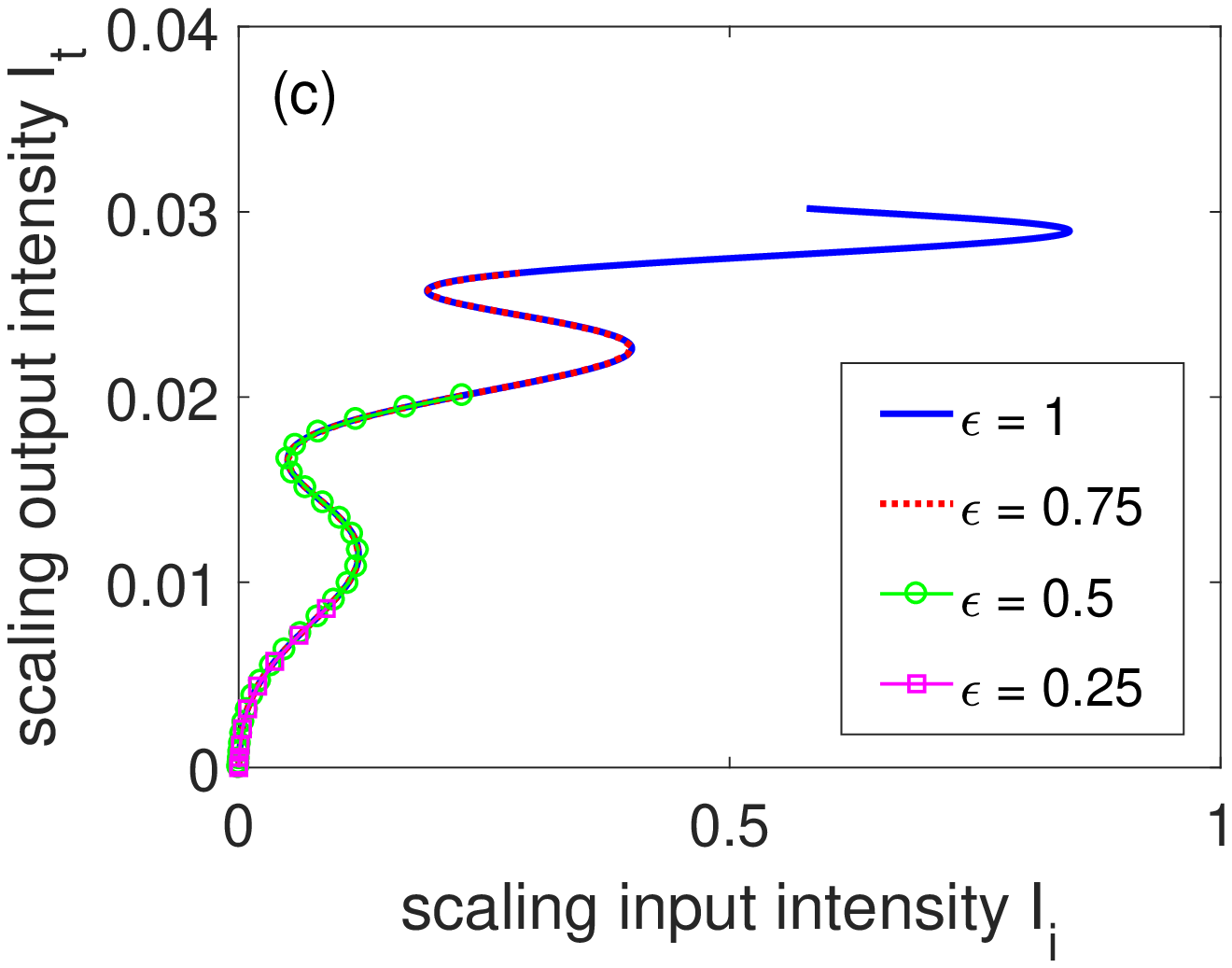} 
\end{center}
\caption{(a) Transmission v.s. output intensity with different optical density $ \alpha $, and the corresponding input-output relation is depicted in (b). The scaling input-output relation is plotted in (c), in which $ \epsilon\equiv n/n_0 $, and $ \rho_0 = 2.4 \times 10^{17} (1/m^3)$. $ \epsilon = 1,~0.75,~0.5 $, and $ 0.25 $ correspond to the four optical densities $ \alpha = 70,~52.5, ~35 $, and $ 17.5 $ in (a) and (b).
\label{FigOD}}
\end{figure}

Finally, we discuss about the influence of optical density of the Rydberg EIT medium.  
Since $ \eta $ is proportional to $ n^2 $, we can expect that the 
output intensity $ I_t $ would be greatly decreased when the optical density $ \alpha $ increases due to large absorption. Nevertheless, the dispersive part is also enhanced at the same time, which forms more bistabilities.  
In Fig.~\ref{FigOD}(a), the relation of transmission versus output intensity is plotted with various optical densities $ \alpha $. 
One can see that the transmission peak shifts to higher $ I_t $ when $ \alpha $ decreases, while we have obtained more bistabilities in
the corresponding input-output relation shown in Fig.~\ref{FigOD}(b). 
The results reflect the fact that the input intensity is smaller for the occurrence of optical bistability when the nonlinearity is larger.
The results described above quantitatively match our physical picture.
Next, we plot the scaling input-output relation in Fig.~\ref{FigOD}(c).
According to the fact of $ \eta \propto n^2 $, we re-scale the input and output intensity $ I_{i,t} \rightarrow \epsilon^2 I_{i,t}$, in which $ \epsilon \equiv n/n_0 $. The four different lines which are plotted with different $ \alpha $ are falling on the same curve. 
Essentially, this rescaling phenomenon shown in Fig.~\ref{FigOc}(b) and  Fig.~\ref{FigOD}(c) can be understood by considering that the output field intensity $ I_t $ is scaled  by the factor $ b $, which is 
proportional to $ n^2 $ and $ \Omega_c^{-3} $.    

\section{iv. conclusion}

In conclusion, we theoretically investigate the optical bistable features of a Rydberg EIT atomic sample in a unidirectional optical ring-cavity. 
In the presence of the van der Walls interactions between the atoms, $ \chi^{(3)} $ nonlinear coefficient of the medium is greatly enhanced, which enables an optical bistable response. A large magnitude of the nonlinear coefficient $\eta$ is not sufficient for a bistable response to appear. This is because, its real and imaginary parts correspond to different phenomena, nonlinear dispersion and absorption, respectively. The nonlinear dispersion supports the bistability, while nonlinear absorption can work against it. Fortunately, the ratio between the nonlinear absorption and dispersion can be adjusted by probe-detuning $ \Delta_p $, which changes the phase of the nonlinear coefficient. That is, absorption can be reduced while dispersive term increases. Interestingly, we observe scaling of the bistable features by the factors of the coupling Rabi frequency $ \Omega_c $ and the atomic optical density $ \alpha $. The nonlinear coefficient also changes with the cavity parameters, such as cavity detuning $ \delta $, and mirror transmission coefficient $ T $, which can be changed in the experiment. We provide an understanding on the effects of the physical  parameters on the bistability features, which can be helpful for the design of optical bistable devices employing Rydberg atomic media.

\section{acknowledgements}

The authors would like to give their acknowledgement to Dr. Julius Ruseckas, in particalr for the discussions on the analytical formula of Rydberg EIT system. 
Also, authors acknowledge National center for Theoretical Sciences for providing computational facilities.
The work is supported by the Ministry of Science and Technology of Taiwan.
MET acknowledges support from TUBITAK 1001 grant no: 117F118 and TUBA-GEBIP 2017.

\appendix*
\section{appendix A}
\renewcommand{\theequation}{A\thesection.\arabic{equation}}
\setcounter{equation}{0}

In appendix A, we will give the detail derivations of Eq.~(\ref{rho}) as well as Eq.~(\ref{kerrFE}) which is the main equation in this paper.
According to Eq.(\ref{s12}, \ref{s13}), we can obtain the steady state by ignoring the time derivatives on left hand side. Thus we have
\begin{eqnarray}
&&\hat{\sigma}_{12} = \dfrac{i}{\Gamma}\left( \Omega_p + \Omega_c^{\ast}\hat{\sigma}_{13}\right), \label{ss_s12}\\
&&\hat{\sigma}_{13} = - \dfrac{\Omega_p\Omega_c + 2i\Gamma \hat{U} \hat{\sigma}_{13}}{\Omega^2} \label{ss_s13} 
\end{eqnarray} 
where $ \Gamma \equiv 2\gamma_{12} - 2i\Delta_p $, and $ \Omega^2 \equiv 2\Gamma\gamma_{13}+\vert\Omega_c\vert^2 $.  
It is clearly to see that Eq.~(\ref{ss_s12}) and Eq.~(\ref{ss_s13}) are  coupled, and they are not final solutions because we still have $ \hat{U} $ which includes $ \hat{\sigma}_{33} $ in Eq.(\ref{ss_s13}). 
By considering the first order atomic operators, we can express 
$ \hat{\sigma}_{33} \simeq \hat{\sigma}_{31}\hat{\sigma}_{13} = \hat{\sigma}_{13}^{\dagger}\hat{\sigma}_{13}$.

With the operator approximation, we can substitute $ \hat{U} $ into Eq.~(\ref{ss_s13}), then we have
\begin{eqnarray}
\hat{\sigma}_{13} = - \dfrac{\Omega_p \Omega_c}{\Omega^2} - \dfrac{2i\Gamma}{\Omega^2}\left[ \int d^3 \textbf{r}' V(\textbf{r}-\textbf{r}')n(\textbf{r}')\hat{\sigma}_{13}^{\dagger}(\textbf{r}')\hat{\sigma}_{13}(\textbf{r}') \right] \hat{\sigma}_{13} \nonumber \\
\label{s13s13} 
\end{eqnarray}   
In Eq.~(\ref{s13s13}), we have $ \hat{\sigma}_{13} $ on both right and left sides. We can do the iteration by re-substituting $ \hat{\sigma}_{13} $ on right hand side again and again. For each time, we will encounter the integration as following.
\begin{eqnarray}
&&\int d^3 \textbf{r}' \int d^3 \textbf{r}''V(\textbf{r}-\textbf{r}')V(\textbf{r}-\textbf{r}'')n(\textbf{r}')n(\textbf{r}'')\nonumber\\ 
&&~~~~~~~~~~~~\times\hat{\sigma}_{13}^{\dagger}(\textbf{r}')\hat{\sigma}_{13}(\textbf{r}')\hat{\sigma}_{13}^{\dagger}(\textbf{r}'')\hat{\sigma}_{13}(\textbf{r}'') \nonumber\\
&&\simeq \int d^3 \textbf{r}' \left[ V(\textbf{r}')\right]^2\left[ n(\textbf{r}')\right]^2\left( \dfrac{V}{N}\right)\hat{\sigma}_{13}^{\dagger}(\textbf{r}')\hat{\sigma}_{13}(\textbf{r}')  \nonumber\\\label{iii}
\end{eqnarray}
where we have used $ \left[ \hat{\sigma}_{13}(\textbf{r}'), \hat{\sigma}_{13}^{\dagger}(\textbf{r}'')\right] = (V/N)\delta(\textbf{r}'-\textbf{r}'')  $.

After the many time iterations by the help of Eq.(\ref{iii}), we can express $ \hat{\sigma}_{13} $ as shown as

\begin{eqnarray}
&&\hat{\sigma}_{13} = - \dfrac{\Omega_p \Omega_c}{\Omega^2} - \dfrac{\Omega_p \Omega_c}{\Omega^2}\dfrac{N}{V} \nonumber\\
&&~~~~~\times\int d^3 \textbf{r}'\sum_{m=1}^\infty 
\left[ -\dfrac{2i\Gamma}{\Omega^2}
 V(\textbf{r}-\textbf{r}')\right]^m \hat{\sigma}_{13}^{\dagger}(\textbf{r}')\hat{\sigma}_{13}(\textbf{r}') \nonumber\\
&& = - \dfrac{\Omega_p \Omega_c}{\Omega^2} +2i\Gamma \dfrac{\Omega_p \Omega_c}{\Omega^2}\dfrac{N}{V}\nonumber\\
&&~~~~~\times\int d^3 \textbf{r}' \dfrac{V(\textbf{r}-\textbf{r}')}{\Omega^2 + 2i\Gamma V(\textbf{r}-\textbf{r}')} \hat{\sigma}_{13}^{\dagger}(\textbf{r}')\hat{\sigma}_{13}(\textbf{r}')\nonumber\\ \label{math}
\end{eqnarray}
From Eq.~(\ref{iii}) to Eq.~(\ref{math}), we have assumed $ n(\textbf{r}) \simeq N/V $. Again, we have $ \hat{\sigma}_{13}^{\dagger}(\textbf{r}')\hat{\sigma}_{13}(\textbf{r}') $ on right hand side, and we try to truncate the iteration by using the lowest order expression of $ \hat{\sigma}_{13} $ which is the first term on right hand side in Eq.~(\ref{math}). 
By doing so, we can see that $ \hat{\sigma}_{13}^{\dagger}(\textbf{r}')\hat{\sigma}_{13}(\textbf{r}') \simeq \vert\Omega_p(\textbf{r}')\vert^2\vert\Omega_c\vert^2/\vert\Omega\vert^2 $.
Re-substituting this expression back to Eq.~(\ref{math}) and using Eq.~(\ref{ss_s12}), we can obtain the formula shown in Eq.~(\ref{rho}).

In our system, $ V(\textbf{r}-\textbf{r}') = C_6/\vert \textbf{r}-\textbf{r}'\vert^6 $, which is vdW potential. After substituting $ V $ into Eq.~(\ref{rho}), we will encounter an integral as shown as follows.
\begin{eqnarray}
&&~~~~\int d^3 \textbf{r}'\dfrac{V(\textbf{r}-\textbf{r}')\vert\Omega_p(\textbf{r}')\vert^2}{\Omega^2 +2i\Gamma V(\textbf{r}-\textbf{r}')} \nonumber\\
&&\simeq \int d^3\textbf{r}' \dfrac{C_6}{\Omega^2\vert\textbf{r} - \textbf{r}'\vert^6 + 2i\Gamma C_6}\cdot\vert\Omega_p(\textbf{r})\vert^2  \nonumber\\
&& \simeq 4\pi\vert\Omega_p(\textbf{r})\vert^2 \int_0^\infty dR\dfrac{R^2 C_6}{\Omega^2 R^6 + 2i\Gamma C_6} \nonumber\\
&& = \dfrac{\pi^2 C_6 \vert C_6\vert^{-1/2}(1-i)}{3\sqrt{\Gamma}\Omega}\vert\Omega_p(\textbf{r})\vert^2 \label{math2}
\end{eqnarray}
in which we have used the approximation $ \Omega_p(\textbf{r}')\simeq \Omega_p(\textbf{r}) $.

Together with Eq.~(\ref{math}, \ref{math2}) and Eq.~(\ref{FE}, \ref{kerrFE}) and the comparison of field propagation equation given by
\begin{eqnarray}
\dfrac{\partial E_p}{\partial \zeta} = i\left( \dfrac{k L \Gamma_2^2}{2}\chi^{(3)}\right) \vert E_p\vert^2 E_p, 
\end{eqnarray}
we can obtain the nonlinear coefficient $ \eta $ given by
\begin{eqnarray}
\eta \simeq \dfrac{\pi^2\alpha}{3}n\left( \dfrac{\Gamma_2}{\vert\Omega_c\vert}\right)^3\sqrt{\dfrac{C_6}{\Gamma_2}} \left( -1+i\right) \left(1-2i\dfrac{\Delta_p}{\Gamma_2} \right)^{-1/2}
\end{eqnarray}
where $ \alpha = n\sigma_{abs}L $ is the absorption cross section of probe field. Using the relation of $ C_6/R_c^6 = 2\vert\Omega_c\vert^2/\Gamma_2$, one can easily to obtain the form of Eq.~(\ref{ita}).

\section{appendix B}
\renewcommand{\theequation}{B\thesection.\arabic{equation}}
\setcounter{equation}{0}

In Appendix B, we will give detail derivations of the analytical solution shown in Eq.~(\ref{Epsol}). 
According to Eq.~(\ref{kerrFE}), we have the field propagation equation with a complex nonlinear coefficient which denotes as $ \eta = a + ib $. Thus Eq.~(\ref{kerrFE}) can be rewritten as
\begin{eqnarray}
\dfrac{\partial E_p}{\partial \zeta} = (-b + ia)\vert E_p\vert^2 E_p \label{Eq_eta_ab}
\end{eqnarray}
From Eq.~(\ref{Eq_eta_ab}), we can clearly to see that the damping term is coming from the coefficient $ b $, at the same time, the coefficient $ a $ arises phase shift. As a result, we can assume that the solution is given by
\begin{eqnarray}
E_p(\zeta) = E_p(0)\exp\left[ -\dfrac{\Lambda(\zeta)}{2} + i\varphi(\zeta)\right], \label{assumption}
\end{eqnarray}  
in which $ \Lambda $ and $ \varphi $ correspond to the damping and phase shift terms. Substituting Eq.~(\ref{assumption}) into Eq.~(\ref{Eq_eta_ab}), we can obtain the following two equations by comparing with real and imaginary parts.
\begin{eqnarray}
&&\dfrac{1}{2}\dfrac{\partial \Lambda}{\partial \zeta} = b \vert E_p(0)\vert^2 e^{-\Lambda}\label{lambda_sol}\\
&&\dfrac{\partial \varphi}{\partial \zeta} = a\vert E_p(0)\vert^2 e^{-\Lambda}\label{phi_sol}
\end{eqnarray}
Solving Eq.~(\ref{lambda_sol}) , we can easily obtain the solution given as 
\begin{eqnarray}
\Lambda(\zeta) = \ln\left( 1+2b\vert E_p(0)\vert^2 \zeta\right). \label{lambda_formula}
\end{eqnarray}
Together with Eq.~(\ref{lambda_formula}) and Eq.~(\ref{phi_sol}), we can solve $ \varphi $ as shown as
\begin{eqnarray}
\varphi(\zeta) = \dfrac{a}{2b}\ln\left( 1+2b\vert E_p(0)\vert^2 \zeta\right) = \dfrac{a}{2b}\Lambda(\zeta). \label{phi_formula}
\end{eqnarray}
Substituting Eq.~(\ref{lambda_formula}) and (\ref{phi_formula}) back into Eq.~(\ref{assumption}), we can obtain Eq.~(\ref{Epsol}). 
It can be found that $ \varphi(1) = a \vert E_p(0)\vert^2 = a I_t/T $ in the absent of nonlinear absorption i.e. $ b\rightarrow 0 $.


\end{document}